\documentstyle[epsfig,11pt]{article}

\newlength{\dinwidth}
\newlength{\dinmargin}
\newcommand{\ba}{\begin{array}}
\newcommand{\ea}{\end{array}}
\newcommand{\bd}{\begin{displaymath}}
\newcommand{\ed}{\end{displaymath}}
\newcommand{\be}{\begin{equation}}
\newcommand{\ee}{\end{equation}}
\newcommand{\bea}{\begin{eqnarray}}
\newcommand{\eea}{\end{eqnarray}}

\begin{document}

\bigskip
\title {Perturbative QCD study of $B_{\small (s)}\to \phi\rho$ decays}
\author{Jing-Wu Li$^a${\footnote {Email:lijw@email.xznu.edu.cn}}, Fu-Yi You$^a${\footnote {Email:youfuyi1981@126.com}}  \\
\it \small $^a$ Department of Physics, Xu Zhou Normal University,
XuZhou 221116, China} \maketitle
\begin{abstract}
We study $B_{\small (s)}\to \phi\rho$ decays in a perturbative QCD
approach based on $k_T$ factorization. In this approach, we
calculate factorizable and non-factorizable contributions, there
are no annihilation contributions due to quark content. We get the
branching ratios and polarization fractions for $B_{\small (s)}\to
\phi\rho$ decays . Our predictions are consistent with the current
experimental data.
\end{abstract}
\bigskip
\section{Introduction}
Exclusive $B$ meson decays, especially $B\to VV$ modes, have
aroused more and more interest both theoretically and
experimentally. Since the first observed charmless $B\to VV$ mode,
the $B\to\phi K^*$ decay\cite{li}, many $B\to VV$decay channels
have been studied in PQCD approach, such as $B\to K^*\rho,
K^*\omega$\cite{lu1}, $B\to K^*K^*$\cite{lu2},
$B_s\to\rho(\omega)K^*$\cite{lu3},
$B\to\rho(\omega)\rho(\omega)$\cite{lu4}, and
$B^0\to\phi\phi$\cite{lu5}.  It offers an excellent place to study
the CP violation and search for new physics hints\cite{sanda}.
Because  the hadronization process is non-perturbative in nature,
the essential problem in handling the decay processes is the
separation of different energy scales, i.e., the factorization
assumption. Many approaches based on the factorization assumption
have been developed, such as the naive factorization\cite{bsw},
the generalized factorization\cite{ali,yh}, the QCD
factorization\cite{bbns}, and the perturbative QCD approach which
is based on $k_T$ factorization\cite{keum,pipi}.

Recently, $B\to \phi{K^*}$ data reveal a large transverse
polarization fraction, which has been considered as a puzzle, many
theoretical efforts have been put to clarify
it\cite{grossman,ygr,alk,wpr,cdp,hn,hnlism,pkdkc}. This suggests
that $B\to VV$ modes must be more complicated than all the other
modes and need to be studied deeply. Motivated by this, we study
another $B\to VV$ mode in the perturbative approach (PQCD) within
the Standard Model. For $B\to \phi\rho$ decay, only penguin
operators contribute and we find that the branching ratio is at
the order of $10^{-9}$. For $B_s^0\to \phi\rho^0$ decay,
current-current operators and  penguin operators can contribute
and we find that the branching ratio is at the order of $10^{-7}$.
The longitudinal polarization predominates over transverse
polarization and its fraction is found to go beyond $70\%$. Our
predictions are consistent with the current experimental values.
We hope that our study will help to resolve the above-mentioned
puzzle a bit.

The remaining part of this paper is organized as follows. In
Sec.II, we calculate analytically the related Feynman diagrams and
present the various decay amplitudes for the decay modes studied.
In Sec.III, we give the numerical analysis for the branching
ratios and polarization fraction of the related decay modes and
compare them with the measured values. The summary and some
discussion are included in the final section.
\section{Theoretical framework and perturbative calculation }
In PQCD approach, the decay amplitude is expressed as the
convolution of the mesons' light-cone wave functions, the hard
scattering kernels and the Wilson coefficients, which stand for
the soft, hard and harder dynamics, respectively. The formalism
can be written as: \vspace{-0.7cm}

\begin{eqnarray}
{\cal M}\sim&&\int dx_1dx_2dx_3b_1db_1b_2db_2b3db_3Tr[C(t)\nonumber\\&&\Phi_B(x_1,b_1)\Phi_{\phi}(x_2,b_2)\Phi_{\rho}(x_3,b_3)\nonumber\\
&&H(x_i,b_i,t)S_t(x_i)e^{-S(t)}]
\end{eqnarray}

where $Tr$ denotes the trace over Dirac and color indices. $C(t)$
is Wilson coefficient of the four-quark operator which results
from the radiative corrections at short distance. The wave
function $\phi_M$ absorbs non-perturbative dynamics of the
process, which is process independent. The hard part $H$ is rather
process-independent and can be calculated in perturbative
approach. The $b_i$ is the conjugate space coordinate of the
transverse momentum, which represents the transverse interval of
the meson. $t$ is the largest energy scale in hard function $H$,
while the jet function $S_t(x_i)$ comes from the resummation of
the double logarithms $\ln^2x_i$, called threshold
resummation\cite{ktl,threshold}, which becomes larger near the
endpoint. The Sudakov form factor $S(t)$ is from the resummation
of double logarithms $\ln^2Qb$ \cite{cheng}.

In this paper, we use the light-cone coordinates to describe the
four-dimensional momentum as $(p^+, p^-, {\bf P}_T)$. We work in
the frame with the $B$ meson at rest, so the meson momentum can be
written as
 \begin{eqnarray}
P_1=\frac{M_B}{\sqrt{2}}(1,1,{\bf 0}_T),\nonumber\\
P_2=\frac{M_B}{\sqrt{2}}(1-r_{\rho}^2,r_{\phi}^2,{\bf
0}_T),\nonumber\\
P_3=\frac{M_B}{\sqrt{2}}(r_{\rho}^2,1-r_{\phi}^2,{\bf 0}_T)
\end{eqnarray}
in which $r_{\rho}, r_{\phi}$ is defined by
$r_{\rho}=M_{\rho}/M_B$ and $r_{\phi}=M_{\phi}/M_B$. $P_1, P_2,
P_3$ refer to $B, \phi, \rho$ respectively. To extract the
helicity amplitudes, we parameterize the following polarization
vectors. The longitudinal polarization must satisfy the
orthogonality and normalization: $\varepsilon_{2L}\cdot P_2=0,
\varepsilon_{3L}\cdot P_3=0$, and
$\varepsilon_{2L}^2=\varepsilon_{3L}^2=-1$. Then we can give the
manifest form as follows: \vspace{-0.6cm}
\begin{center}
\begin{eqnarray}
\varepsilon_{2L}&=\frac{1}{\sqrt{2}r_{\phi}}(1-r_{\phi}^2,-r_{\phi}^2,{\bf
0}_T)\nonumber\\
\varepsilon_{3L}&=\frac{1}{\sqrt{2}r_{\rho}}(-r_{\rho}^2,1-r_{\rho}^2,{\bf
0}_T)
\end{eqnarray}
\end{center}

As to the transverse polarization vectors, we can choose the
simple form: \vspace{-0.6cm}
\begin{center}
\begin{eqnarray}
\varepsilon_{2T}&=\frac{1}{\sqrt{2}}(0,0,{\bf 1}_T)\nonumber\\
\varepsilon_{3T}&=\frac{1}{\sqrt{3}}(0,0,{\bf 1}_T)
\end{eqnarray}
\end{center}

The decay width for these channel is:
\begin{equation}
\Gamma=\frac{G_F^2{\bf
|P_c|}}{16{\pi}M_B^2}\sum\limits_{\sigma=L,T}{\cal
M}^{\sigma\dagger}{\cal M}^{\sigma}
\end{equation}
where $\bf |P_c|$ is the three-dimensional momentum of the final
state meson, and ${\bf
|P_c|}=\frac{M_B}{2}(1-r_{\rho}^2-r_{\phi}^2)$. The subscript
$\sigma$ denotes the helicity states of the two vector mesons with
$L(T)$ standing for the longitudinal(transverse) component. As
discussed in Ref.\cite{li}, the amplitude ${\cal M}^{\sigma}$ is
decomposed into
\begin{eqnarray}
{\cal M}^{\sigma}=M_B^2{\cal M}_L+M_B^2{\cal
M}_N\varepsilon_2^*(\sigma=T)\cdot\varepsilon_3^*(\sigma=T)\nonumber\\+i{\cal
M}_T\varepsilon_{\mu\nu\rho\sigma}\varepsilon_2^{\mu*}\varepsilon_3^{\nu*}P_2^{\rho}P_3^{\sigma}
\end{eqnarray}

We can define the longitudinal $H_0$, transverse $H_{\pm}$
helicity amplitudes
\begin{equation}
H_0=M_B^2{\cal M}_L, H_{\pm}=M_B^2{\cal M}_N\mp
M_{\phi}M_{\rho}\sqrt{r^2-1}{\cal M}_T
\end{equation}
where $r=P_2\cdot P_3/(M_{\phi}M_{\rho})$. And we can deduce that
they satisfy the relation
\begin{equation}
\sum\limits_{\sigma=L,T}{\cal M}^{\sigma\dagger}{\cal
M}^{\sigma}=|H_0|^2+|H_+|^2+|H_-|^2
\end{equation}

There is another set of definition of helicity amplitudes
\vspace{-0.7cm}
\begin{center}
\begin{eqnarray}
A_0=-\xi M_B^2{\cal M}_L,\nonumber\\
A_{\|}=\xi\sqrt{2}M_B^2{\cal M}_N,\nonumber\\
A_{\perp}=\xi M_{\phi}M_{\rho}\sqrt{2(r^2-1)}{\cal M}_T
\end{eqnarray}
\end{center}
with the normalization factor $\xi=\sqrt{G_F^2P_c/(16\pi
M_B^2\Gamma)}$. These helicity amplitudes satisfy the relation,
\begin{equation}
|A_0|^2+|A_{\|}|^2+|A_{\perp}|^2=1
\end{equation}
where the notation $A_0, A_{\|}, A_{\perp}$ denote the
longitudinal, parallel, and perpendicular polarization amplitudes.

Our next task is to calculate the matrix elements ${\cal M}_L,
{\cal M}_N$ and ${\cal M}_T$ of the operators in the weak
Hamiltonian with PQCD approach. We have to use the mesons'
light-cone wave functions, they are universal for all decay
channels. We employ the following wave functions as in other PQCD
calculations \cite{lu1,lu2,lu3}.
\begin{eqnarray}
& &\frac{1}{\sqrt{2N_c}}(\not P_1+M_B)\gamma_5\Phi_{B}(x,b)\;,
\nonumber\\
& &\frac{1}{\sqrt{2N_c}}[M_\phi\not \epsilon_2(L)\Phi_\phi(x)
+\not\epsilon_2(L)\not P_2 \Phi_{\phi}^{t}(x)+M_\phi
I\Phi_\phi^s(x)]\;,
\nonumber\\
& &\frac{1}{\sqrt{2N_c}} [M_\phi\not \epsilon_2(T)\Phi_\phi^v(x)+
\not\epsilon_2(T)\not P_2\Phi_\phi^T(x) +\frac{M_\phi}{P_2\cdot
n_-}
i\epsilon_{\mu\nu\rho\sigma}\gamma_5\gamma^\mu\epsilon_2^\nu(T)
P_2^\rho n_-^\sigma \Phi_\phi^a(x)]\;,
\nonumber\\
& &\frac{1}{\sqrt{2N_c}}[M_{\rho}\not \epsilon_3(L)\Phi_{\rho}(x)
+\not\epsilon_3(L)\not P_3 \Phi_{\rho}^{t}(x)+M_{\rho}
I\Phi_{\rho}^s(x)]\;,
\nonumber\\
& &\frac{1}{\sqrt{2N_c}} [M_{\rho}\not
\epsilon_3(T)\Phi_{\rho}^v(x)+ \not\epsilon_3(T)\not
P_3\Phi_{\rho}^T(x) +\frac{M_{K^*}}{P_3\cdot n_+}
i\epsilon_{\mu\nu\rho\sigma}\gamma_5\gamma^\mu\epsilon_3^\nu(T)
P_3^\rho n_+^\sigma \Phi_{\rho}^a(x)]\;. \nonumber
\end{eqnarray}
where $n_+=(1,0,{\bf 0}_T)$ and  $n_-=(0,1,{\bf 0}_T)$ are
dimensionless vectors on the light cone.  $x$ is the momentum
fraction.
\subsection{$B\to\phi\rho$ decays}
 The effective Hamiltonian for the process$B\to\phi\rho$ is given
as \cite{Buras}
\begin{eqnarray}
{\cal H}_{eff}=\frac{G_F}{\sqrt{2}}\{V_u
[C_1(\mu)O_1^{u}(\mu)+C_2(\mu)O^{u}_2(\mu)]
\nonumber\\-V_t\sum_{i=3}^{10}C_i(\mu)O_i^{(q)}(\mu)\},
\end{eqnarray}
where  $V_u=V_{ud}^*V_{ub}$ , $V_t=V_{td}^*V_{tb}$ , $C_i(\mu)$
are the Wilson coefficients, and the operators are
\begin{eqnarray}
\nonumber &&O^{u}_1=(\bar{d}_iu_j)_{V-A}(\bar{u}_jb_i)_{V-A},\,
 O^{u}_2=(\bar{d}_iu_i)_{V-A}(\bar{u}_jb_j)_{V-A}, \nonumber\\
\nonumber
&&O_3=(\bar{d}_ib_i)_{V-A}\sum\limits_q(\bar{q}_jq_j)_{V-A},\,
O_4=(\bar{d}_ib_j)_{V-A}\sum\limits_q(\bar{q}_jq_i)_{V-A}, \nonumber\\
\nonumber &&
O_5=(\bar{d}_ib_i)_{V-A}\sum\limits_q(\bar{q}_jq_j)_{V+A},\,
 O_6=(\bar{d}_ib_j)_{V-A}\sum\limits_q(\bar{q}_jq_i)_{V+A},\nonumber\\ \nonumber
 &&O_7=\frac{3}{2}(\bar{d}_ib_i)_{V-A}\sum\limits_qe_q(\bar{q}_jq_j)_{V+A},\,
 O_8=\frac{3}{2}(\bar{d}_ib_j)_{V-A}\sum\limits_qe_q(\bar{q}_jq_i)_{V+A},\nonumber
 \\ &&
 O_9=\frac{3}{2}(\bar{d}_ib_i)_{V-A}\sum\limits_qe_q(\bar{q}_jq_j)_{V-A},\,
 O_{10}=\frac{3}{2}(\bar{d}_ib_j)_{V-A}\sum\limits_qe_q(\bar{q}_jq_i)_{V-A}.
\end{eqnarray}
Here i and j stand for $SU(3)$ color indices. The sum over $q$
runs over the quark fields that are active at the scale
$\mu=O(m_b)$, i.e., $(q\in\{u,d,s,c,b\})$. From the effective
Hamiltonian, we can see that the current-current operators have no
contribution. For factorizable diagrams, all the penguin operators
contribute, but for the non-factorizable diagrams only the
operators $O_4,O_6,O_8,O_{10}$ can contribute because of the color
structure. The leading order diagrams for these decays are shown
in Fig.1. We first calculate the usual factorization diagrams
(a)and (b). \hspace{-0.4cm}
\begin{flushleft}
\begin{eqnarray}
&F_{Le}=8\pi C_FM_B^2\int_{0}^{1}dx_1dx_3\int_{0}^{\infty}b_1db_1b_3db_3\Phi_B(x_1,b_1)\nonumber\\
&\times\{\left[(1+x_3)\Phi_{\rho}(x_3)+r_{\rho}(1-2x_3)(\Phi_{\rho}^t(x_3)+\Phi_{\rho}^a(x_3))\right]\nonumber\\
&\times E_e(t_e^{(1)})h_e(x_1,x_3,b_1,b_3)\nonumber\\
&+2r_{\rho}\Phi_{\rho}^s(x_3)E_e(t_e^{(2)})h_e(x_3,x_1,b_3,b_1)\}
\end{eqnarray}
\end{flushleft}
\vspace{-0.95cm}
\begin{flushleft}
\begin{eqnarray}
&F_{Ne}=8\pi C_FM_B^2\int_{0}^{1}dx_1dx_3\int_{0}^{\infty}b_1db_1b_3db_3\Phi_B(x_1,b_1)\nonumber\\
&\times r_{\phi}\{\left[\Phi_{\rho}^T(x_3)+2r_{\rho}\Phi_{\rho}^v(x_3)+r_{\rho}x_3(\Phi_{\rho}^v(x_3)-\Phi_{\rho}^a(x_3))\right]\nonumber\\
&\times E_e(t_e^{(1)})h_e(x_1,x_3,b_1,b_3)\nonumber\\
&+r_{\rho}[\Phi_{\rho}^v(x_3)+\Phi_{\rho}^a(x_3)]E_e(t_e^{(2)})h_e(x_3,x_1,b_3,b_1)\}
\end{eqnarray}
\end{flushleft}
\vspace{-0.95cm}
\begin{flushleft}
\begin{eqnarray}
&F_{Te}=16\pi C_FM_B^2\int_{0}^{1}dx_1dx_3\int_{0}^{\infty}b_1db_1b_3db_3\Phi_B(x_1,b_1)\nonumber\\
&\times
r_{\phi}\{\left[\Phi_{\rho}^T(x_3)+2r_{\rho}\Phi_{\rho}^a(x_3)-r_{\rho}x_3(\Phi_{\rho}^v(x_3)-\Phi_{\rho}^a(x_3))\right]\nonumber\\
&\times E_e(t_e^{(1)})h_e(x_1,x_3,b_1,b_3)\nonumber\\
&+r_{\rho}[\Phi_{\rho}^v(x_3)+\Phi_{\rho}^a(x_3)]E_e(t_e^{(2)})h_e(x_3,x_1,b_3,b_1)\}
\end{eqnarray}
\end{flushleft}
where $C_F={4\over 3}$ is a color factor. The function $h_e$,
including the jet function $S_t(x_3)$(threshold resummation for
non-factorizable diagrams is weaker and negligible ), is the same
as the $h_e$ in \cite{li} . The factors $E(t)$ contain the
evolution from the $W$ boson mass to the hard scales $t$ in the
Wilson coefficients $a(t)$, and from t to the factorization scale
$1/b$ in the Sudakov factors $S(t)$:
\begin{equation}
\label{aet} E_e(t)=\alpha_s(t)a_e(t)S_B(t)S_{\rho}(t)
\end{equation}

The Wilson coefficients $a_e(t)$ in Eq.\ref{aet} are given by
\begin{equation}
a_e(t)=C_3+\frac{C_4}{3}+C_5+\frac{C_6}{3}-\frac{C_7}{2}-\frac{C_8}{6}-\frac{C_9}{2}-\frac{C_{10}}{6}
\end{equation}

For the non-factorizable diagrams (c) and (d), all the three meson
wave functions are involved and the amplitudes ${\cal
M}_{He}={\cal M}_{He4}+{\cal M}_{He6}$ are written as
\vspace{-0.4cm}
\begin{flushleft}
\begin{eqnarray}
&{\cal M}_{Le4}=16\pi
C_FM_B^2\sqrt{2N_c}\int_{0}^{1}dx_1dx_2dx_3\int_{0}^{\infty}b_1db_1b_2db_2\Phi_B(x_1,b_1)\nonumber\\
&\times\{\Phi_{\phi}(x_2)\left[-(x_2+x_3)\Phi_{\rho}(x_3)+r_{\rho}x_3(\Phi_{\rho}^t(x_3)+\Phi_{\rho}^s(x_3))\right]\nonumber\\
&\times E_{e4}(t_d^{(1)})h_d^{(1)}(x_1,x_2,x_3,b_1,b_2)\nonumber\\
&+\Phi_{\phi}(x_2)\left[(1-x_2)\Phi_{\rho}(x_3)+r_{\rho}x_3(\Phi_{\rho}^t(x_3)-\Phi_{\rho}^s(x_3))\right]\nonumber\\
&\times E_{e4}(t_d^{(2)})h_d^{(2)}(x_1,x_2,x_3,b_1,b_2)
\end{eqnarray}
\end{flushleft}
\vspace{-0.95cm}
\begin{flushleft}
\begin{eqnarray}
&{\cal M}_{Ne4}=16\pi C_FM_B^2\sqrt{2N_c}\int_{0}^{1}dx_1dx_2dx_3\int_{0}^{\infty}b_1db_1b_2db_2\Phi_B(x_1,b_1)\nonumber\\
&\times r_{\phi}\{[x_2(\Phi_{\phi}^v(x_2)+\Phi{\phi}^a(x_2))\Phi_{\rho}^T(x_3)\nonumber\\
&-2r_{\rho}(x_2+x_3)(\Phi_{\phi}^v(x_2)\Phi_{\rho}^v(x_3)+\Phi_{\phi}^a(x_2)\Phi_{\rho}^a(x_3))]\nonumber\\
&\times E_{e4}(t_d^{(1)})h_d^{(1)}(x_1,x_2,x_3,b_1,b_2)\nonumber\\
&+(1-x_2)(\Phi_{\phi}^v(x_2)+\Phi_{\phi}^a(x_2))\Phi_{\rho}^T(x_3)\nonumber\\
&\times E_{e4}(t_d^{(2)})h_d^{(2)}(x_1,x_2,x_3,b_1,b_2)\}
\end{eqnarray}
\end{flushleft}
\vspace{-0.95cm}
\begin{flushleft}
\begin{eqnarray}
&{\cal M}_{Te4}=32\pi C_FM_B^2\sqrt{2N_c}\int_{0}^{1}dx_1dx_2dx_3\int_{0}^{\infty}b_1db_1b_2db_2\Phi_B(x_1,b_1)\nonumber\\
&\times r_{\phi}\{[x_2(\Phi_{\phi}^v(x_2)+\Phi{\phi}^a(x_2))\Phi_{\rho}^T(x_3)\nonumber\\
&-2r_{\rho}(x_2+x_3)(\Phi_{\phi}^v(x_2)\Phi_{\rho}^a(x_3)+\Phi_{\phi}^a(x_2)\Phi_{\rho}^v(x_3))]\nonumber\\
&\times E_{e4}(t_d^{(1)})h_d^{(1)}(x_1,x_2,x_3,b_1,b_2)\nonumber\\
&+(1-x_2)(\Phi_{\phi}^v(x_2)+\Phi_{\phi}^a(x_2))\Phi_{\rho}^T(x_3)\nonumber\\
&\times E_{e4}(t_d^{(2)})h_d^{(2)}(x_1,x_2,x_3,b_1,b_2)\}
\end{eqnarray}
\end{flushleft}
\vspace{-0.95cm}
\begin{flushleft}
\begin{eqnarray}
&{\cal M}_{Le6}=-16\pi
C_FM_B^2\sqrt{2N_c}\int_{0}^{1}dx_1dx_2dx_3\int_{0}^{\infty}b_1db_1b_2db_2\Phi_B(x_1,b_1)\nonumber\\
&\times\Phi_{\phi}(x_2)\{\left[x_2\Phi_{\rho}(x_3)+r_{\rho}x_3(\Phi_{\rho}^t(x_3)-\Phi_{\rho}^s(x_3))\right]\nonumber\\
&\times E_{e6}(t_d^{(1)})h_d^{(1)}(x_1,x_2,x_3,b_1,b_2)\nonumber\\
&+\left[-(1-x_2+x_3)\Phi_{\rho}(x_3)+r_{\rho}x_3(\Phi_{\rho}^t+\Phi_{\rho}^s(x_3))\right]\nonumber\\
&\times E_{e6}(t_d^{(2)})h_d^{(2)}(x_1,x_2,x_3,b_1,b_2)
\end{eqnarray}
\end{flushleft}
\vspace{-0.95cm}
\begin{flushleft}
\begin{eqnarray}
&{\cal M}_{Ne6}=-16\pi C_FM_B^2\sqrt{2N_c}\int_{0}^{1}dx_1dx_2dx_3\int_{0}^{\infty}b_1db_1b_2db_2\Phi_B(x_1,b_1)\nonumber\\
&\times r_{\phi}\{x_2(\Phi_{\phi}^v(x_2)-\Phi{\phi}^a(x_2))\Phi_{\rho}^T(x_3)E_{e6}(t_d^{(1)})h_d^{(1)}(x_1,x_2,x_3,b_1,_b2)\nonumber\\
&+[(1-x_2)(\Phi_{\phi}^v(x_2)-\Phi_{\phi}^a(x_2))\Phi_{\rho}^T(x_3)\nonumber\\
&-2r_{\rho}(1-x_2+x_3)(\Phi_{\phi}^v(x_2)\Phi_{\rho}^v(x_3)-\Phi_{\phi}^a(x_2)\Phi_{\rho}^a(x_3))]\nonumber\\
&\times E_{e6}(t_d^{(2)})h_d^{(2)}(x_1,x_2,x_3,b_1,b_2)\}
\end{eqnarray}
\end{flushleft}
\vspace{-0.95cm}
\begin{flushleft}
\begin{eqnarray}
&{\cal M}_{Te6}=-32\pi C_FM_B^2\sqrt{2N_c}\int_{0}^{1}dx_1dx_2dx_3\int_{0}^{\infty}b_1db_1b_2db_2\Phi_B(x_1,b_1)\nonumber\\
&\times r_{\phi}\{x_2(\Phi_{\phi}^v(x_2)-\Phi{\phi}^a(x_2))\Phi_{\rho}^T(x_3)E_{e6}(t_d^{(1)})h_d^{(1)}(x_1,x_2,x_3,b_1,_b2)\nonumber\\
&+[(1-x_2)(\Phi_{\phi}^v(x_2)-\Phi_{\phi}^a(x_2))\Phi_{\rho}^T(x_3)\nonumber\\
&-2r_{\rho}(1-x_2+x_3)(\Phi_{\phi}^v(x_2)\Phi_{\rho}^a(x_3)-\Phi_{\phi}^a(x_2)\Phi_{\rho}^v(x_3))]\nonumber\\
&\times E_{e6}(t_d^{(2)})h_d^{(2)}(x_1,x_2,x_3,b_1,b_2)\}
\end{eqnarray}
\end{flushleft}

The evolution factors are given by
\begin{equation}
E_{ei}(t)=\alpha_s(t)a_i(t)S(t)|_{b_3=b_1}
\end{equation}
with the Sudakov factor $S=S_BS_{\phi}S_{\rho}$. The Wilson
coefficients $a$ appearing in the above formulas are
\vspace{-0.5cm}
\begin{center}
\begin{eqnarray}
&a_4=\frac{C_4}{3}-\frac{C_{10}}{6},\nonumber\\
&a_6=\frac{C_6}{3}-\frac{C_8}{6}
\end{eqnarray}
\end{center}

The amplitudes for$B^+\to \phi\rho^+$ are written as
\begin{equation}
{\cal M}_H=f_{\phi}V_t^*F_{He}+V_t^*{\cal M}_{He}
\end{equation}
where $F_{He}$ denotes the factorizable contributions and ${\cal
M}_{He}$ the non-factorizable contributions. For the other two
decay channels of $B\to\phi\rho$, the amplitudes are the
following: for $B^-\to\phi\rho^-$,
\begin{equation}
\bar{\cal M}_H=f_{\phi}V_tF_{He}+V_t\bar{\cal M}_{He}
\end{equation}

and for $B^0\to\phi\rho^0$,
\begin{equation}
-\sqrt{2}{\cal M}_H^0=V_t^*f_{\phi}F_{He}+V_t^*{\cal M}_{He}
\end{equation}

\subsection{$B_s^0\to\phi\rho^0$ decay}

For the process $B_s^0\to \phi\rho^0$ , it is a $b\to
 s $ transition and we use the effective Hamiltonian
\cite{Buras}
\begin{eqnarray}
\label{heff} {\cal H}_{{\it eff}} = \frac{G_{F}} {\sqrt{2}} \,
\left[ V_{ub} V_{us}^*  (C_1 O_1^u + C_2 O_2^u \right) - V_{tb}
V_{ts}^* \, (\sum_{i=3}^{10} C_{i} \, O_i^{(q)} ) ] \quad .
\end{eqnarray}

We specify below the operators in ${\cal H}_{{\it eff}}$ for $b
\to s$:
\begin{eqnarray}
\nonumber &&O_1^{u}=(\bar{s}_iu_j)_{V-A}(\bar{u}_jb_i)_{V-A},\,
O_2^{u}=(\bar{s}_iu_i)_{V-A}(\bar{u}_jb_j)_{V-A},\nonumber
\end{eqnarray}
\begin{eqnarray}
\nonumber
&&O_3=(\bar{s}_ib_i)_{V-A}\sum\limits_q(\bar{q}_jq_j)_{V-A},\,
O_4=(\bar{s}_ib_j)_{V-A}\sum\limits_q(\bar{q}_jq_i)_{V-A},\nonumber \\
\nonumber &&
O_5=(\bar{s}_ib_i)_{V-A}\sum\limits_q(\bar{q}_jq_j)_{V+A},\,
 O_6=(\bar{s}_ib_j)_{V-A}\sum\limits_q(\bar{q}_jq_i)_{V+A},\nonumber\\ \nonumber
 &&O_7=\frac{3}{2}(\bar{s}_ib_i)_{V-A}\sum\limits_qe_q(\bar{q}_jq_j)_{V+A},\,
 O_8=\frac{3}{2}(\bar{s}_ib_j)_{V-A}\sum\limits_qe_q(\bar{q}_jq_i)_{V+A},\nonumber
 \\ &&
 O_9=\frac{3}{2}(\bar{s}_ib_i)_{V-A}\sum\limits_qe_q(\bar{q}_jq_j)_{V-A},\,
 O_{10}=\frac{3}{2}(\bar{s}_ib_j)_{V-A}\sum\limits_qe_q(\bar{q}_jq_i)_{V-A}.
\end{eqnarray}
From the effective Hamiltonian we can see that the current-current
operators and penguin operators can contribute. The leading order
diagrams for the decay are shown in Fig.2, the amplitude for
$B_s^0\to \rho^0\phi$ mode is
\begin{equation}
-\sqrt{2}{\cal
M}_H^{\prime}=V_u^*f_{\rho}F_{He}^{\prime}+V_u^*{\cal
M}_{He}^{\prime}-V_t^*f_{\rho}F_{He}^{P\prime}-V_t^*{\cal
M}_{He}^{P\prime}
\end{equation}
where ${\cal M}_{He}^{P\prime}={\cal M}_{He4}^{P\prime}+{\cal
M}_{He6}^{P\prime}$. Here $V_u=V_{ub}V_{us}^*$,
$V_t=V_{tb}V_{ts}^*$ and amplitude for the corresponding CP
conjugate model is written as
\begin{equation}
-\sqrt{2}\bar{{\cal
M}}_H^{\prime}=V_uf_{\rho}F_{He}^{\prime}+V_u{\cal
M}_{He}^{\prime}-V_tf_{\rho}F_{He}^{P\prime}-V_t{\cal
M}_{He}^{P\prime}
\end{equation}
Next we calculate the hard part in PQCD approach. For the
factorizable diagram (e) and (f), we have
\begin{flushleft}
\begin{eqnarray}
&F_{Le}^{\prime}=8\pi C_FM_{B_s}^2\int_{0}^{1}dx_1dx_2\int_{0}^{\infty}b_1db_1b_2db_2\Phi_{B_s}(x_1,b_1)\nonumber\\
&\times\{\left[(1+x_2)\Phi_{\phi}(x_2)+r_{\phi}^{\prime}(1-2x_2)(\Phi_{\phi}^t(x_2)+\Phi_{\phi}^a(x_2))\right]\nonumber\\
&\times E_e^{\prime}(t_e^{(1)})h_e(x_1,x_2,b_1,b_2)\nonumber\\
&+2r_{\phi}^{\prime}\Phi_{\phi}^s(x_2)E_e^{\prime}(t_e^{(2)})h_e(x_2,x_1,b_2,b_1)\}
\end{eqnarray}
\end{flushleft}
\vspace{-0.95cm}
\begin{flushleft}
\begin{eqnarray}
&F_{Ne}^{\prime}=8\pi C_FM_{B_s}^2\int_{0}^{1}dx_1dx_2\int_{0}^{\infty}b_1db_1b_2db_2\Phi_{B_s}(x_1,b_1)\nonumber\\
&\times r_{\rho}^{\prime}\{\left[\Phi_{\phi}^T(x_2)+2r_{\phi}^{\prime}\Phi_{\rho}^v(x_2)+r_{\phi}x_2(\Phi_{\phi}^v(x_2)-\Phi_{\phi}^a(x_2))\right]\nonumber\\
&\times E_e^{\prime}(t_e^{(1)})h_e(x_1,x_2,b_1,b_2)\nonumber\\
&+r_{\phi}^{\prime}[\Phi_{\phi}^v(x_2)+\Phi_{\phi}^a(x_2)]E_e^{\prime}(t_e^{(2)})h_e(x_2,x_1,b_2,b_1)\}
\end{eqnarray}
\end{flushleft}
\vspace{-0.95cm}
\begin{flushleft}
\begin{eqnarray}
&F_{Te}^{\prime}=16\pi C_FM_{B_s}^2\int_{0}^{1}dx_1dx_3\int_{0}^{\infty}b_1db_1b_2db_2\Phi_{B_s}(x_1,b_1)\nonumber\\
&\times
r_{\rho}^{\prime}\{\left[\Phi_{\phi}^T(x_2)+2r_{\phi}^{\prime}\Phi_{\phi}^a(x_2)-r_{\phi}^{\prime}x_2(\Phi_{\phi}^v(x_2)-\Phi_{\phi}^a(x_2))\right]\nonumber\\
&\times E_e^{\prime}(t_e^{(1)})h_e(x_1,x_2,b_1,b_2)\nonumber\\
&+r_{\phi}^{\prime}[\Phi_{\rho}^v(x_2)+\Phi_{\phi}^a(x_2)]E_e(t_e^{(2)})h_e(x_2,x_1,b_2,b_1)\}
\end{eqnarray}
\end{flushleft}
with $r_{\phi}^{\prime}, r_{\rho}^{\prime}$ equal to $r_{\phi},
r_{\rho}$ except for $M_B$ replaced by $M_{B_s}$. The expression
for $F_{He}^{P\prime}$ are the same as $F_{He}^{\prime}$ but with
WIlson coefficients $a$ replaced by $a^P$. As before, the factor
$E^{\prime}(t)$ are given by
\begin{equation}
E_e^{(P)\prime}(t)=\alpha_s a_e^{(P)}(t)S_B(t)S_{\phi}(t)
\end{equation}
where the Wilson coefficients are the following
\begin{center}
\begin{eqnarray}
a=C_1+2/3C_2,\nonumber\\
a^P=3/2C_7+1/2C_8+3/2cC_9+1/2C_{10}.
\end{eqnarray}
\end{center}
\vspace{-0.3cm}For non-factorizable diagram (g) and (h), we find
that
\begin{flushleft}
\begin{eqnarray}
&{\cal M}_{Le4}^{\prime}=16\pi
C_FM_{B_s}^2\sqrt{2N_c}\int_{0}^{1}dx_1dx_2dx_3\int_{0}^{\infty}b_1db_1b_3db_3\nonumber\\&\Phi_{B_s}(x_1,b_1)
\times\{\Phi_{\rho}(x_3)[-(x_2+x_3)\Phi_{\phi}(x_2)+r_{\phi}^{\prime}x_2(\Phi_{\phi}^t(x_2)\nonumber\\&+\Phi_{\phi}^s(x_2))]\times E_{e4}^{(q)\prime}(t_d^{(1)})h_d^{(1)}(x_1,x_2,x_3,b_1,b_3)\nonumber\\
&+\Phi_{\rho}(x_3)\left[(1-x_3)\Phi_{\phi}(x_2)+r_{\phi}^{\prime}x_2(\Phi_{\phi}^t(x_2)-\Phi_{\phi}^s(x_2))\right]\nonumber\\
&\times
E_{e4}^{(q)\prime}(t_d^{(2)})h_d^{(2)}(x_1,x_2,x_3,b_1,b_3)\}
\end{eqnarray}
\end{flushleft}
\vspace{-0.95cm}
\begin{flushleft}
\begin{eqnarray}
&{\cal M}_{Le6}^{\prime}=-16\pi
C_FM_B^2\sqrt{2N_c}\int_{0}^{1}dx_1dx_2dx_3\int_{0}^{\infty}b_1db_1b_3db_3\nonumber\\
&\Phi_{B_s}(x_1,b_1)\times\Phi_{\rho}(x_3)\{[x_3\Phi_{\phi}(x_2)+r_{\phi}^{\prime}x_2(\Phi^t_{\phi}(x_2)-\nonumber\\
&\Phi_{\phi}^s(x_2))]\times E_{e6}^{(q)\prime}(t_d^{(1)})h_d^{(1)}(x_1,x_2,x_3,b_1,b_3)\nonumber\\
&+\left[-(1-x_3+x_2)\Phi_{\phi}(x_2)+r_{\phi}^{\prime}x_2(\Phi_{\phi}^t(x_2)+\Phi_{\phi}^s(x_2))\right]\nonumber\\
&\times
E_{e6}^{(q)\prime}(t_d^{(2)})h_d^{(2)}(x_1,x_2,x_3,b_1,b_3)\
\end{eqnarray}
\end{flushleft}

\begin{flushleft}
\begin{eqnarray}
&{\cal M}_{Ne4}^{q\prime}=16\pi C_FM_{B_s}^2\sqrt{2N_c}\int_{0}^{1}dx_1dx_2dx_3\int_{0}^{\infty}b_1db_1b_3db_3\nonumber\\
&\Phi_{B_s}(x_1,b_1)\times r_{\rho}^{\prime}\{[x_3(\Phi_{\rho}^v(x_3)+\Phi_{\rho}^a(x_3))\Phi_{\phi}^T(x_2)\nonumber\\
&-2r_{\phi}^{\prime}(x_2+x_3)(\Phi_{\rho}^v(x_3)\Phi_{\phi}^v(x_2)+\Phi_{\rho}^a(x_3)\Phi_{\phi}^a(x_2))]\nonumber\\
&\times E_{e4}^{(q)\prime}(t_d^{(1)})h_d^{(1)}(x_1,x_2,x_3,b_1,b_3)\nonumber\\
&+(1-x_3)(\Phi_{\rho}^v(x_3)+\Phi_{\rho}^a(x_3))\Phi_{\phi}^T(x_2)\nonumber\\
&\times
E_{e4}^{(q)\prime}(t_d^{(2)})h_d^{(2)}(x_1,x_2,x_3,b_1,b_3)\}
\end{eqnarray}
\end{flushleft}

\begin{flushleft}
\begin{eqnarray}
&{\cal M}_{Ne6}^{\prime}=-16\pi
C_FM_B^2\sqrt{2N_c}\int_{0}^{1}dx_1dx_2dx_3\int_{0}^{\infty}b_1db_1\nonumber\\
&b_3db_3\Phi_{B_s}(x_1,b_1)\times r_{\rho}^{\prime}\{[x_3\left(\Phi_{\rho}^v(x_3)-\Phi_{\rho}^a(x_3)\right)\nonumber\\
&\Phi_{\phi}^T(x_2)]\times E_{e6}^{(q)\prime}(t_d^{(1)})h_d^{(1)}(x_1,x_2,x_3,b_1,b_3)\nonumber\\
&+[(1-x_3)\left(\Phi_{\rho}^v(x_3)-\Phi_{\rho}^a(x_3)\right)\Phi_{\phi}^T(x_2)-2r_{\phi}^{\prime}\nonumber\\
&(1-x_3+x_2)
\left(\Phi_{\rho}^v(x_3)\Phi_{\phi}^v(x_2)-\Phi_{\rho}^a(x_3)\Phi_{\phi}^a(x_2)\right)]\nonumber\\
&\times
E_{e6}^{(q)\prime}(t_d^{(2)})h_d^{(2)}(x_1,x_2,x_3,b_1,b_3)\}
\end{eqnarray}
\end{flushleft}

\begin{flushleft}
\begin{eqnarray}
&{\cal M}_{Te4}^{\prime}=32\pi C_FM_B^2\sqrt{2N_c}\int_{0}^{1}dx_1dx_2dx_3\int_{0}^{\infty}b_1db_1b_3\nonumber\\
&db_3\Phi_{B_s}(x_1,b_1)\times r_{\rho}^{\prime}\{[x_3(\Phi_{\rho}^v(x_3)+\Phi_{\rho}^a(x_3))\Phi_{\phi}^T(x_2)\nonumber\\
&-2r_{\phi}^{\prime}(x_2+x_3)(\Phi_{\rho}^v(x_3)\Phi_{\phi}^a(x_2)+\Phi_{\rho}^a(x_3\Phi_{\phi}^v(x_2))]\nonumber\\
&\times E_{e}^{(q)\prime}(t_d^{(1)})h_d^{(1)}(x_1,x_2,x_3,b_1,b_3)\nonumber\\
&+(1-x_3)(\Phi_{\rho}^v(x_3)+\Phi_{\rho}^a(x_3))\Phi_{\phi}^T(x_2)\nonumber\\
&\times
E_{e}^{(q)\prime}(t_d^{(2)})h_d^{(2)}(x_1,x_2,x_3,b_1,b_3)\}
\end{eqnarray}
\end{flushleft}
\begin{flushleft}
\begin{eqnarray}
&{\cal M}_{Te6}^{\prime}=-16\pi
C_FM_B^2\sqrt{2N_c}\int_{0}^{1}dx_1dx_2dx_3\int_{0}^{\infty}b_1db_1\nonumber\\
&b_3db_3\Phi_{B_s}(x_1,b_1)\times r_{\rho}^{\prime}\{x_3[\Phi_{\rho}^v(x_3)-\Phi_{\rho}^a(x_3)]\Phi_{\phi}^T(x_2)\nonumber\\
&\times E_{e}^{(q)\prime}(t_d^{(1)})h_d^{(1)}(x_1,x_2,x_3,b_1,b_3)\nonumber\\
&+[(1-x_3)\left(\Phi_{\rho}^v(x_3)-\Phi_{\rho}^a(x_3)\right)\Phi_{\phi}^T(x_2)-2r_{\phi}^{\prime}\nonumber\\
&(1-x_3+x_2)
\left(\Phi_{\rho}^v(x_3)\Phi_{\phi}^a(x_2)-\Phi_{\rho}^a(x_3)\Phi_{\phi}^v(x_2)\right)]\nonumber\\
&\times
E_{e}^{(q)\prime}(t_d^{(2)})h_d^{(2)}(x_1,x_2,x_3,b_1,b_3)\}
\end{eqnarray}
\end{flushleft}
The evolution factors are given by
\begin{equation}
E_{ei}^{(q)\prime}(t)=\alpha_s(t)a_{ei}^{(q)\prime}(t)S(t)|_{b_2=b_1}
\end{equation}
with the Sudakov factor $S=S_{B_s}S_{\phi}S_{\rho}$ and the Wilson
coefficients are given by
\begin{eqnarray}
a_1^{\prime}=C_2/N_c,\nonumber\\
a_4^{\prime}=3/2C_{10}/N_c,\nonumber\\
a_6^{\prime}=3/2C_{8}/N_C.
\end{eqnarray}

\subsection{Numerical analysis}
The parameters used in our calculations are: the Fermi coupling
constant $G_F=1.16639\times 10^{-5}GeV^{-2}$, the meson masses
$M_B=5.28GeV$, $M_{B_S}=5.37GeV$, $M_{\rho}=0.77GeV$,
$M_{\phi}=1.02GeV$, the decay constant $f_{\rho}=0.205GeV$,
$f_{\rho}^T=0.155GeV$, $f_{\phi}=0.237GeV$, $f_{\phi}^T=0.220GeV$,
the central value of the CKM matrix elements $\gamma=60^{\circ}$,
$|V_{td}|=0.0075,|V_{tb}|=0.9992$,
$|V_{ub}|=0.0047,|V_{us}|=0.2196$ and the meson
lifetime$\tau_B=1.65ps,\tau_{B_s}=1.461ps$\cite{particle}.

Using the above parameters, we  get the branching ratios and
helicity amplitudes of $B_{\small (s)}\to \phi\rho$ decays (the
helicity amplitudes are in Table 1)
\begin{eqnarray}
B_r(B^\pm\to\phi\rho^\mp)=4.1\times 10^{-9},\nonumber\\
B_r(B^0\to\phi\rho^0)=1.9\times 10^{-9},\nonumber\\
B_r(B_s^0\to \phi\rho^0)=3.09\times 10^{-7},\nonumber\\
B_r(\bar{B_s}^0\to \phi\rho^0)=3.60\times 10^{-7}
\end{eqnarray}
Compared  with the averaged results  of QCDF\cite{Hai}
\begin{eqnarray}
BR(B^-\to \rho^-\phi)=5.5\times 10^{-9},\nonumber\\
 BR(\bar{B}^0\to\phi\rho^0)=2.5\times 10^{-9}.
\end{eqnarray}
or those from naive factorization \cite{guo}
\begin{eqnarray}
BR(\bar{B}_s^0\to \rho^0\phi)=2.92\times 10^{-7}.
\end{eqnarray}
our predictions for $B\to \phi\rho$ are consistent with those of
QCDF, the predictions  for $\bar{B}_s^0\to \rho^0\phi$ are
consistent with the result of  naive factorization, because the
nonfactorization effects in $\bar{B}_s^0\to \rho^0\phi$  are
little.

Presently, only the experimental upper limits are available at the
90\% confidence level \cite{particle},
\begin{eqnarray}
B_r(B^+\to\phi\rho^+)<1.6\times 10^{-5},\nonumber\\
B_r(B^0\to\phi\rho^0)<1.3\times 10^{-5},\nonumber\\
B_r(B_s^0\to \phi\rho^0)<6.17\times 10^{-4}.
\end{eqnarray}

Obviously, our results are consistent with the data. Our
predictions will be tested by the oncoming measurements.

For $B\to\phi\rho$ decays, only penguin operators contribute, so
there is no direct CP violation  in the $B\to\phi\rho$ decays.

For $B_s^0\to \phi\rho^0$ decays, the CP asymmetry is time
dependent
\begin{equation}
A_{CP} (t) \simeq A_{CP}^{dir} \cos (\Delta mt) + a_{\epsilon +
\epsilon '} \sin (\Delta m t),
\end{equation}
where $\Delta m$ is the mass difference of the two  mass
eigenstates of neutral $B_{s}$ mesons.

The direct CP violation parameter is defined
\begin{eqnarray}
A_{CP}^{dir} &=& \frac{|{\cal M}|^2 -|\overline{\cal M}|^2}{
|{\cal M}|^2 +|\overline{\cal M}|^2}.
\end{eqnarray}
The direct CP violation parameter we can get in $B_s^0\to
\phi\rho^0$ is
\begin{eqnarray}
A_{CP}^{dir}(B_s^0\to \phi\rho^0) &=-8.0\%.
\end{eqnarray}

From Table 1, we can find the longitudinal fraction in $B_s^0\to
\phi\rho^0$ decays go beyond $70\%$, but the longitudinal
fractions in $B\to \phi\rho$ decays are  very small, which is
similar to $B^0\to \rho^0\rho^0$ decay mode\cite{c0601019}. In
$B\to \phi\rho$ decays, $O_{1,2}$in ${\cal H}_{{\it eff}}$ don't
contribute via factorizable diagrams, the penguin operators
contributing via factorizable diagrams are color suppressed, so
that the nonfactorizable effects are the same order as the
factorizable ones, which cause the $B\to \phi\rho$ decays  not to
be  factorization dominated, besides the above reason, the
nonfactorizable longitudinal amplitude is opposite in sign to that
of the factorizable part, therefore the longitudinal amplitude
gets a large cancellation between the factorizable effects and the
non factorizable parts such that $|A_0|^2$ is reduced much.

According to Ref.\cite{yang}, we can get the $B_{(s)}\to\rho,\phi$
vector meson transition form factors ,
\begin{eqnarray}
&B\to\rho,  &V(0)=0.303,
A_0(0)=0.308,\nonumber\\& &A_1(0)=0.233,A_2(0)=0.208,\nonumber\\
&B_s\to\phi, &V(0)=0.430,A_0(0)=0.363,\nonumber\\&
&A_1(0)=0.304,A_2(0)=0.276,\nonumber
\end{eqnarray}
which are  consistent with the results with light cone sum rules
\cite{ball}.
\begin{center}
 \begin{table}[tbh]
 \begin{center}
\begin{tabular}{|c|c|c|c|}\hline
Channel & $|A_0|^2$ & $|A_{\|}|^2$ &$|A_{perp}|^2$\\
\hline $B^\pm\to\phi\rho^\mp$ & 0.14 & 0.41 &0.45\\
\hline $B^0({\bar B}^0)\to\phi\rho^0$&0.14 & 0.41 &0.45\\
\hline
$B_s^0\to\phi\rho^0$&0.78&0.12&0.10\\
\hline
${\bar B}_S^0\to\phi\rho^0$&0.83&0.09&0.08\\
\hline
\end{tabular}
\vspace{0.3cm}\caption{Branching ratios and helicity amplitudes }
\end{center}
\end{table}
\end{center}
\section{Summary}
In this paper, we calculate the branching ratios and polarization
fractions of $B\to \phi\rho$ and $B_s^0\to \phi\rho^0$ decays in
perturbative QCD approach, the predicted branching ratios are
compared with the experimental data and results obtained with
other approaches.
 CP parameters in $B_s^0\to \phi\rho^0$ are given in our paper. we
compared  with the experimental values, our results are consistent
with the current experimental data. \vskip 0.3cm

We thank Dong-Sheng Du and Mao-Zhi Yang for helpful discussions
and communications.

\begin{appendix}
\section{wave functions}
The $\phi$ and $\rho$ distribution amplitudes up to twist 3 are
given by
\begin{eqnarray}
&&\Phi_{\phi}(x)=\frac{3f_{\phi}}{\sqrt{2N_c}}x(1-x),\nonumber\\
&&\Phi^t_{\phi}(x)=\frac{f_{\phi}^T}{2\sqrt{2N_c}}\{3(1-2x)^2
+1.68C_4^{\frac{1}{2}}(1-2x)
+0.69[1+(1-2x)\ln{\frac{x}{1-x}}]\},\nonumber\\
\nonumber
 &&\Phi^s_{\phi}(x)=\frac{f_{\phi}^T}{4\sqrt{2N_c}}[3(1-2x)
 (4.5-11.2x+11.2x^2)+1.38\ln{\frac{x}{1-x}}],\nonumber\\
&&\Phi^T_{\phi}(x)=\frac{3f_{\phi}^T}{2\sqrt{2N_c}}x(1-x)
\left[1+0.2C_4^{\frac{3}{2}}(1-2x)
\right],\nonumber\\
&&\Phi^v_{\phi}(x)=\frac{f_{\phi}^T}{2\sqrt{2N_c}}\{\frac{3}{4}
[1+(1-2x)^2]+0.24[3(1-2x)^2-1]
+0.96C_4^{\frac{1}{2}}(1-2x)\},\nonumber\\
&&\Phi^a_{\phi}(x)=\frac{3f_{\phi}^T}{4\sqrt{2N_c}}(1-2x)[1+0.93(10x^2-10x+1)].\nonumber\\
\end{eqnarray}
with the Gegenbauer polynomials,
\begin{eqnarray}
&&C_2^{\frac{1}{2}}(\xi)=\frac{1}{2}(3\xi^2-1),\nonumber\\
&&C_4^{\frac{1}{2}}(\xi)=\frac{1}{8}(35\xi^4-30\xi^2+3),\nonumber\\
&&C_2^{\frac{3}{2}}(\xi)=\frac{3}{2}(5\xi^2-1).
\end{eqnarray}

For $\rho$ meson , its Lorentz structures are similar to $\phi$
meson and the distribution amplitudes are given by
\begin{eqnarray}
&&\Phi_\rho(x)=\frac{3f_\rho}{\sqrt{2N_c}} x(1-x)\left[1+
0.18C_2^{3/2}(1-2x)\right]\;,
\label{pwr}\nonumber\\
&&\Phi_{\rho}^t(x)=\frac{f^T_{\rho}}{2\sqrt{2N_c}}
\{3(1-2x)^2+0.3(1-2x)^2[5(1-2x)^2-3]
+0.21[3-30(1-2x)^2+35(1-2x)^4]\}\;,
\label{pwt}\nonumber\\
&&\Phi_{\rho}^s(x) =\frac{3f_\rho^T}{2\sqrt{2N_{c}}}
(1-2x)[1+0.76(10x^2-10x+1)]\;,
\label{pws}\nonumber\\
 &&\Phi_\rho^T(x)=\frac{3f_\rho^T}{\sqrt{2N_c}}
x(1-x)\left[1+ 0.2C_2^{3/2}(1-2x)\right]\;,
\label{pwft}\nonumber\\
&&\Phi_{\rho}^v(x)=\frac{f_{\rho}}{2\sqrt{2N_c}}
\bigg\{\frac{3}{4}[1+(1-2x)^2]+0.24[3(1-2x)^2-1]
+0.12[3-30(1-2x)^2+35(1-2x)^4]\bigg\}\;,
\label{pwv}\nonumber\\
&&\Phi_{\rho}^a(x) =\frac{3f_\rho}{4\sqrt{2N_{c}}}
(1-2x)\left[1+0.93(10x^2-10x+1)\right]\;. \label{pwa}\nonumber
\end{eqnarray}

For the amplitudes of $B$ and$B_s$, meson, we employ the following
distribution amplitudes:
\begin{equation}
\Phi_B(x,b)=N_Bx^2(1-x)^2\exp[-\frac{M_B^2x^2}{2\omega_b^2}-\frac{1}{2}(\omega_bb)^2]
\end{equation}
which satisfies the normalization
\begin{equation}
\int_0^1\Phi_B(x)dx=\frac{f_B}{2\sqrt{2N_c}}\;. \label{dco}
\end{equation}
We choose $N_B=91.78\mbox{4GeV}$, $\omega_B=0.4\mbox{GeV}$. Things
for $B_s$ are similar if we ignore $SU(3)$ symmetry breaking
effect. As discussed in
Ref. \cite{lu3}, we choose $\omega_{B_s}=0.50$GeV. \\
\end{appendix}

\begin{figure}[h]
\begin{center}
\psfig{file=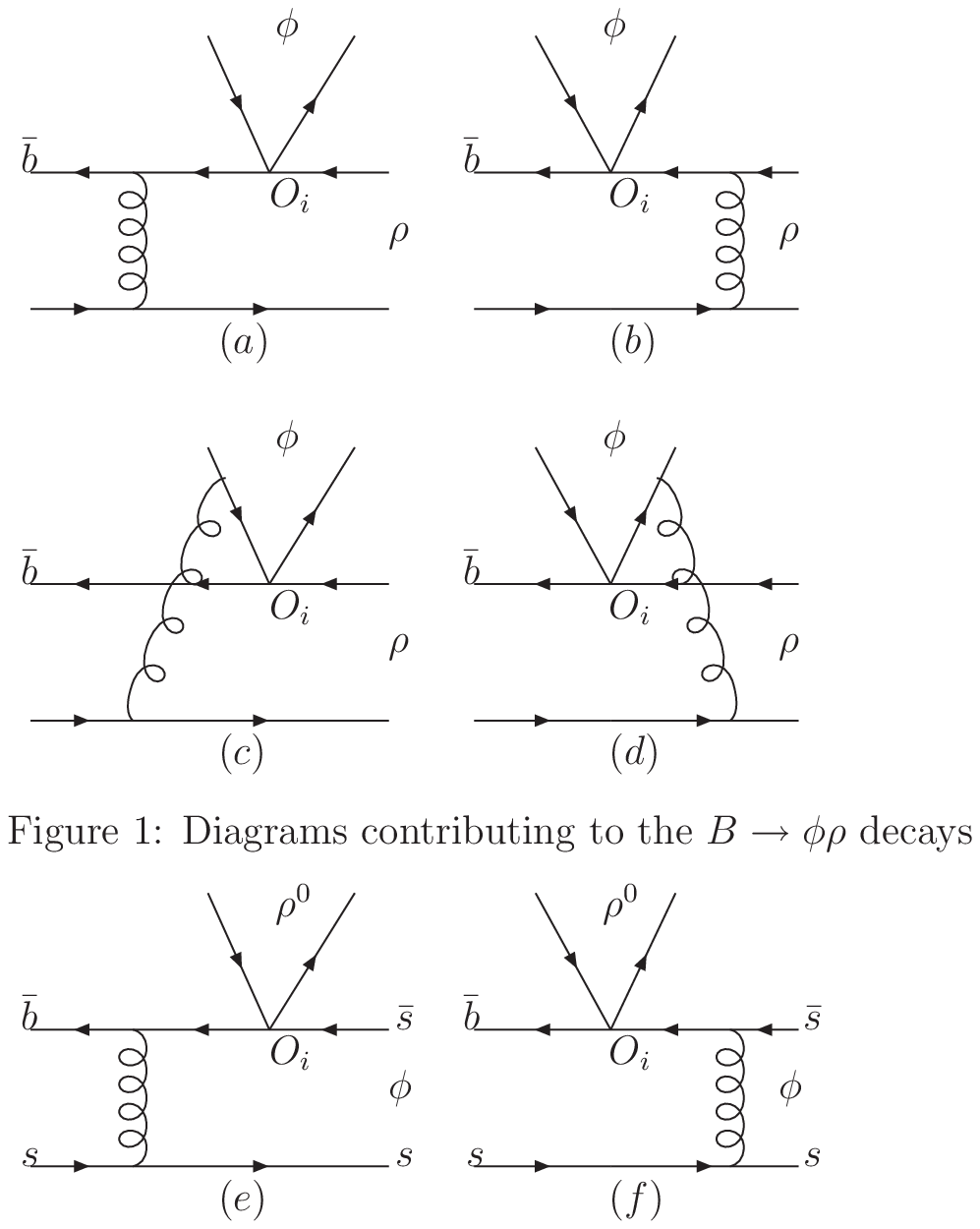,height=26cm,width=18cm}
\end{center}
\end{figure}
\hspace{10pt}

\end{document}